\documentclass{article}

\pdfoutput=1
\usepackage{authblk}
\usepackage[margin=1in]{geometry}
\usepackage[utf8]{inputenc}
\usepackage{graphicx}
\usepackage{amsmath}
\usepackage{amsfonts}
\usepackage{color, colortbl}
\usepackage[colorlinks = true,
            linkcolor = blue,
            urlcolor  = blue,
            citecolor = blue]{hyperref}

\usepackage[mathlines]{lineno}
\usepackage[round]{natbib}

\usepackage[T1]{fontenc}
\usepackage[utf8]{inputenc}
\usepackage{mathptmx}
\usepackage{multicol}

\usepackage[mathscr]{eucal}

\title{The Overfitted Brain: Dreams evolved to assist generalization}
\author[1]{Erik Hoel\thanks{erik.hoel@tufts.edu}}
\affil[1]{Allen Discovery Center, Tufts University, Medford, MA, USA}

\begin{document}

\maketitle

\begin{abstract}

Understanding of the evolved biological function of sleep has advanced considerably in the past decade. However, no equivalent understanding of dreams has emerged. Contemporary neuroscientific theories generally view dreams as epiphenomena, and the few proposals for their biological function are contradicted by the phenomenology of dreams themselves. Now, the recent advent of deep neural networks (DNNs) has finally provided the novel conceptual framework within which to understand the evolved function of dreams. Notably, all DNNs face the issue of overfitting as they learn, which is when performance on one data set increases but the network's performance fails to generalize (often measured by the divergence of performance on training vs. testing data sets). This ubiquitous problem in DNNs is often solved by modelers via “noise injections” in the form of noisy or corrupted inputs. The goal of this paper is to argue that the brain faces a similar challenge of overfitting, and that nightly dreams evolved to combat the brain’s overfitting during its daily learning. That is, dreams are a biological mechanism for increasing generalizability via the creation of corrupted sensory inputs from stochastic activity across the hierarchy of neural structures. Sleep loss, specifically dream loss, leads to an overfitted brain that can still memorize and learn but fails to generalize appropriately. Herein this "overfitted brain hypothesis" is explicitly developed and then compared and contrasted with existing contemporary neuroscientific theories of dreams. Existing evidence for the hypothesis is surveyed within both neuroscience and deep learning, and a set of testable predictions are put forward that can be pursued both in vivo and in silico.
\end{abstract}

\begin{multicols}{2}

\section{\label{sec:introduction}Introduction}

During the Covid-19 pandemic of 2020, many of those in isolation reported an increase in the vividness and frequency of their dreams \citep{weaver_2020}, even leading {\#}pandemicdreams to trend on Twitter. Yet dreaming is so little understood there can be only speculative answers to the \textit{why} behind this widespread change in dream behavior. This is despite the fact that humans spend hours every night dreaming and that dream deprivation is highly damaging to animals \citep{albert1970behavioral}; indeed, dreaming is homeostatically regulated in that there even appears to be a "dream drive" \citep{dement1960effect}. Additionally, dreaming is conserved across many species \citep{siegel1999evolution}, indicating an essential evolved purpose \citep{cirelli2008sleep}. Yet finding a biological function for dreams themselves has evaded resolution. The "null theory" that dreams and even sleep itself are adaptive only in that they prevent organisms from moving during periods which they are not adapted to, such as ensuring that diurnal animals are inactive during night, is still taken seriously \citep{siegel2011sleep}. This is despite the fact that around 50-70\% of the time subjects report having a dream upon a sudden waking during sleep, with more dreams being reported later on in the night \citep{foulkes1962dream, stickgold2001brain}. Reports from those who keep regular dream journals, or who spend time each morning recollecting dreams, indicates that preserving a memory of dreams increases significantly with practice \citep{robb2018we}. All this hints that individuals may regularly underestimate how much time they actually spending dreaming during sleep.

Few contemporary theories appropriately account for the phenomenology of dreams and their sparse, hallucinatory, and narrative contents. What is the purpose of this strange state? The fact that sleep overall has some relationship to learning was known even by the roman orator Quintilian \citep{duff2014secret}. Yet, as will be discussed in detail in section \ref{sec:current_theories_REM}, contemporary neuroscientific theories which relate dreaming to memory storage, memory replay, or emotional processing, still view dreams themselves as epiphenomena.

This lack of viable theories about why animals dream stands in contrast with how much is known about sleep physiology and its stages \citep{lee2012neuromodulation}. As originally discovered by lesion studies and later supported by genetic knockout studies, the sleep state is brought about by a far-reaching set of subcortical neuromodulatory systems, with no one system being necessary, indicating redundancy in how the waking state is sustained \citep{jones2005waking}. In general this multifaceted arousal system is excitatory during wake in that it has the greatest firing, and becomes more quiescent to bring about sleep, although this is not true for all such systems, especially those that establish REM.

A classic signature of sleep are slow waves, which are waves of activity that traverse the cortex, which can be identified when the dominate frequency of EEG is less than 1 Hz. In this state, the cortex become bistable, oscillating between periods of intense firing and periods often referred to as "down states" wherein neurons are silent. In general it should be noted that there is a spectrum wherein sometimes a brain region is experiencing slow wave sleep and this is not synchronized with other regions \citep{nir2011regional}. This is despite the fact that sleep itself is traditionally broken down into NREM sleep and REM sleep, with REM sleep being more associated with dreaming than NREM sleep. Yet there is evidence that dreaming occurs regularly throughout the night, across different sleep stages \citep{oudiette2012dreaming}, although it is rarest in the "deepest" stage of NREM, stage 3, wherein surface EEG reflects low-frequency cortical slow-waves. Recent neuroimaging and sudden-waking experiments have demonstrated that all sleep stages can have dreams, which are the result of localized wake-like firing \citep{siclari2017neural}. On average high-frequency EEG signals in posterior areas of the brain were most correlated with reports on waking. Despite the ubiquity of dreams, it is still the case that REM is most strongly associated with dreaming, over 80-90\% of the time in some awakening studies, althought it should be noted again that sleep involves a spectrum wherein it is difficult to find any stage at any time of the night that does not contain any dreams at all \citep{cipolli2017beyond}. While early on in the night dreams can present themselves as more "thought-like" and simple, later on in the night, particularly during REM, dreams can become incredibly complex with a fully-developed narrative structure \citep{siclari2016sleep}.

As far as is currently known, dreaming of this latter sort is a brain-wide state where the brain is experiencing a single narrative or event, which is supported by the activation of the default-mode network during dreaming \citep{domhoff2011neural}. Consider, for instance, the evidence from sudden-waking experiments that higher frequency activity in the frontal lobe predicted emotional affect within the subject's dream \citep{sikka2019eeg}. Or consider the evidence that during lucid dreaming activity is similar to waking movements in the sensorimotor cortex \citep{dresler2011dreamed}. 

Despite the nuances and redundancies of the cortical systems in play, it still makes sense to view the change from dreamless sleep to dreaming as occurring via a brain-wide neuromodulatory system that regulates level of consciousness, such as by the increase in firing of acetylcholine-containing neurons during REM. Neuromodulatory systems also create the conditions of muscle atonia during dreaming, without which dreams can be acted out by the body during sleep, a dangerous parasomnia called REM sleep behavior disorder \citep{schenck2002rem}.

What is the overall evolved purpose or function of sleep? The evidence of distinct physiological states brought about by neuromodulation suggests answering this requires identifying multiple functions, particularly for dreamless sleep vs. dreaming. Therefore, we should expect dreams to play an important part in the evolved role of sleep. Across the tree of life sleep as a whole is highly conserved; most mammals spend somewhere between 4-20 hours sleeping \citep{joiner2016unraveling}. There is even evidence that \textit{C. elegans} sleeps \citep{trojanowski2016call}. In the past two decades there has been significant progress when it comes to understanding the evolved function of sleep as a whole, although this has not been true for dreams themselves.

First, a novel discovery has led to a clear purpose for at least one aspect of sleep. This was the discovery of the brain's gymphatic system, showing that sleep involves the brain-wide flushing of metabolites with cerebral spinal fluid \citep{xie2013sleep}. This led to the theory that sleep, especially during slow wave activity, had the goal of waste clearance and this is at least partly behind sleep's restorative aspect. Glymphatic activity in the form of this flushing is low in waking but high during both sleep and while under certain types of anesthesia. In sleep and also under anesthesia the greatest amount of flushing occurs during slow wave sleep when low-frequency delta power dominates the EEG \citep{hablitz2019increased}, indicating that it may be anti-correlated with dreaming, although this has not been explicitly established.

Another important theory of the purpose of sleep is the Synaptic Homeostasis Hypothesis (SHY) \citep{tononi2003sleep}. According to SHY, daily learning leads to net synaptic potentiation across the brain, which, if left unchecked, would lead to a saturation of synaptic weight and a cessation of learning \citep{tononi2006sleep}. SHY hypothesizes that slow waves trigger a brain-wide down-scaling of synaptic weights. This indiscriminate down-scaling ensures that the relative weights of synapses are kept proportional while removing the risk of saturation. SHY has served as a model neuroscientific theory in that it has generated a number of new empirical findings \citep{bushey2011sleep}. At the same time, it has also triggered fecund debate and investigation \citep{frank2012erasing}. Traditionally, SHY is more associated with slow wave sleep than with the high-frequencies that indicate dreaming. It also faces the problem that even a global down-scaling of synapses will not keep the pattern of synaptic weights the same, since neurons are nonlinear mechanisms. As an example, a net down-scaling of an artificial neural network with ReLU activation functions would not affect its function; in the case of sigmoid activation functions, it would significantly impact function, and in an unknown way. 

The overall evidence indicates that sleep can be broken into two parts: during deep dreamless sleep metabolic clearance and cellular repair occurs, and some form of unknown contribution to improvements in performance and learning on tasks \citep{cao2020unraveling}. Ultimately, this purpose of the dreaming phase or aspect of sleep lacks hypotheses as explicit and clear as those for slow wave sleep. 

In order to offer forward a distinct theory of the purpose of dreams, this paper outlines the idea that the brains of animals are constantly in danger of \textit{overfitting}, which is the lack of generalizability that occurs in a deep neural network when its learning is based too much on one particular data set, and that dreams help mitigate this ubiquitous issue. This is the \textit{Overfitted Brian Hypothesis} (OBH). The goal of this paper is to formally fill-out the OBH by investigating the evidence that the brain fits to a data set composed of the statistically self-similar daily experiences of the organism, while nightly dreams improve the generality and robustness of an animal's representations, cognition, and perceptual systems, by generating data far outside the organism's daily "training set" in a warped or corrupted way \citep{Hoel_2019}. This idea is supported by the idea that stochasticity (such as corrupted or sparse inputs) is critical in machine learning \citep{sabuncu2020intelligence}. As will be discussed, the OBH fits with known biological understanding and data, matches better with dream phenomenology than other theories, draws various antecedents and similar approaches under one specific roof, and additionally has roots in common practices in deep learning. It makes unique predictions that can be tested both via computational modeling and also in vivo.

\section{\label{sec:current_theories_REM}Contemporary theories of dreams}

A hypothesis for the evolved purpose of sleep must outline a clear and distinct function from other aspects of sleep. It must also explain how dreams present themselves, that is, the phenomenology of dream experience. Specifically, it must explain why dream phenomenology is different than wake phenomenology. Consider three phenomenological properties unique to dreams. First, the \textit{sparseness} of dreams in that they are generally less vivid than waking life and contain less sensory and conceptual information, i.e., less detail. Second, the \textit{hallucinatory} quality of dreams in that they are generally unusual in some way, i.e., not the simple repetition of daily events or specific memories. This includes the fact that in dreams events and concepts often exist outside of normally strict categories (a person becomes another person, a house a spaceship, etc). Third, the \textit{narrative} property of dreams, in that dreams in adult humans are generally sequences of events ordered such that they form a narrative, albeit a fabulist one. As we will see it is not in spite of these properties that dreams serve their evolved purpose, but because of them.

This section explores existent theories of dreams, the supporting evidence (or lack thereof) and how they fail to integrate well with, or explain, dream phenomenology.

\subsection{\label{sec:emotional_regulation}Dreams are for emotional regulation}

The idea that dreams are important for emotional health is a descendant of Freudian theories of psycholanalysis \citep{freud2013interpretation}. While Freud's theories of dreams as expressions of taboo frustration are discredited, there is still a historical association between dreams as expressions of, or important for, emotional processing.

The specific proposals for how dreams impact emotional regulation involve hypotheses like that dreams are somehow for fear extinction \citep{levin2009nightmares}. Such hypotheses reason that dreams might act somewhat like cognitive behavioral therapy treatment for phobias, wherein they provide a safe space for "rehearsals" toward fearful things in order to make them less frightful \citep{scarpelli2019functional}. Yet there is no evidence that the fears of nightmares are the kind of irrational fears faced by those with phobias, nor that fears toward nightmarish events in general should be attenuated, as fear is evolutionarily quite useful. 

Another kind of theory is that dreams act as a kind of "emotional thermostat" in order to regulate emotions \citep{cartwright2005dreaming}. From dream journals there is some evidence that more emotional dreams predict better recovery from disorders like depression \citep{cartwright2006relation}, although sample size for this sort of research is prohibitively small throughout oneirology. From neuroimaging there is evidence that emotional processing centers like the amygdala show greater activity during REM even than during wake \citep{hobson1998dream}, although the role of the amygdala ranges widely from emotions to rewards to motivations \citep{janak2015circuits}. There is some evidence that changes in REM sleep indicate mood disorders \citep{kupfer1976rem}. However, this isn't unique to only REM sleep, as NREM sleep is also changed or reduced in mood disorders \citep{gillin1979successful}, and many cognitive disorders show sleep problems in general \citep{peterson2006sleep}. Sleep deprivation does appear to lead to emotional issues such as a lack of emotional inhibition and also irritability \citep{gruber2014interplay}. But such failures of appropriate function holds true across many cognitive processes following sleep deprivation, including executive function, which would affect emotional regulation \citep{killgore2010effects}.

Overall, the hypothesis that dreams are for resolving emotional conflicts specifically does not have overwhelming empirical evidence. It is also not supported by the phenomenology of dreams, which, at least in general, are not emphatically emotional. Indeed, emotionally neutral dreams are common. Overall there appears to be a slight bias to dreams to have negative affect \citep{merritt1994emotion}, although this may simply be that dreams which have high emotional valence are memorable (and it is worth noting that in studying dream reports joy/elation was the next most common to anxiety/fear). Given the evidence it seems likely that whatever the evolved purpose of dreams is, its function can affect emotions, but there is no strong evidence that dreaming has evolved specifically for emotional regulation.

\subsection{\label{sec:consolidation}Dreams are for memory consolidation}

Perhaps the leading contemporary theory is that dreams somehow involve memory consolidation and storage, often via a proposed form of memory replay \citep{wamsley2014dreaming}. The dominant metaphor for this theory of consolidation is that of the computer: memories need to be "stored" somewhere in the brain, like storing a computer file on a hard drive, and therefore there must be a storage process. This viewpoint is held by much of traditional cognitive neuroscience, wherein the goal of the brain is to "store" memories as veridically as possible \citep{marcus2009kluge}. The theory of memory consolidation is that this storage process occurs during dreams, or alternatively that dreams, by accessing previously stored memories, strengthen them, or that somehow dreams are a byproduct of integrating new memories with older ones.

There is a significant line of research that draws from this theory, including many neuroimaging studies, and a full review of the literature would be beyond the scope of this paper \citep{vorster2015sleep}. However, there is also debate. Specifically the consolidation hypothesis is both very broad and rarely meant to specify just dreams rather than sleep in general \citep{siegel2001rem}. For example, there is evidence that learning a new task leads to a greater activation during both REM \citep{maquet2005human} and slow wave sleep \citep{peigneux2004spatial} in the task-relevant cortical areas. This is true even when comparing a wake/sleep condition versus a control condition without sleep but over the same time, which has found that BOLD activity increased in associated brain regions with the task \citep{debas2010brain}. But these sorts of neuroimaging studies are not very specific, since increased activation in relevant areas does not actually mean storage, nor replay, nor integration with existing memories. Indeed, they could be interpreted just as easily for evidence of the OBH (see Section \ref{sec:neuro_evidence}). 

A significant line of direct evidence for the consolidation theory comes in the form of "replay" of memories during sleep, a specific hypothesis with a clear thesis and standards of evidence. Replay was originally discovered in the hippocampus of rats \citep{wilson1994reactivation, skaggs1996replay}, although the original analysis was again for slow wave sleep, not dreaming specifically. Indeed, the same statistically-increase in correlated neurons that counts as "replay" occurs during quiet wakefulness, indicating it has nothing specific to do with dreams \citep{kudrimoti1999reactivation}. In general, if two neurons potentiate at the same time and from the same cause, they are more likely to be correlated in the sense of increased firing in the future, regardless of whether they are replaying anything specific, a view supported by the finding that "replay" events are much faster in terms of their firing \citep{euston2007fast}.

There are a number of significant issues with the specific hypotheses that dreams are replaying memories \citep{wamsley2014dreaming}. First, there is a lacking strong theoretical argument as to why offline replay of episodic memories would assist memorization, rather than introducing errors, since ground truth is absent offline. Indeed, neuroscience has shown that re-accessing memories generally changes them, rather than enforces them \citep{duvarci2004characterization}.  

Additionally, replay is unlikely to be the purpose of dreams, since, as previously discussed, based on the most detailed studies on dream reports after awakenings, dreams are normally unconnected with the day's events \citep{nielsen2005memory}. Except in cases of PTSD, dreams don't repeat specific memories, and those that do are considered pathological; for instance, closely after Hurricane Andrew, the only hurricane-related dreams, even from a sample of the population from the hardest hit area, were from those already diagnosed with PTSD \citep{david1997dreams}. Overall, it appears that less than 1-2\% of dreams reports have anything to do with episodic memories \citep{stickgold2001sleep}. And this is only when similarity is taken loosely (e.g., involving the same people despite having totally different events occur, or a task but much more hallucinatory or unrealistic). Indeed, there is significant evidence that episodic memory and dreams are dissociated \citep{fosse2003dreaming}. While there is behavioral evidence that repetitive daily tasks, like having subjects play Tetris for significant periods of time, can lead to Tetris-inspired dreams, such inculcated images or sequences do not represent replay in that they are not veridical repeats of previous games, being more hallucinatory and sparse. Moreover, dreams triggered by such repetitive games appear even in patients completely lacking all memory, those with clinically diagnosed amnesia \citep{stickgold2000replaying}. It can take several days for repetitive tasks to show up in dreams, a form of "dream lag," and almost always these tasks appear in partial forms that are, again, only loosely similar \citep{blagrove2011assessing}. Overall, the behavioral evidence suggests that dreams are not replays of memories or waking events.

Additionally, well-controlled neuroimaging experiments show little evidence for exact sequence replication and strong evidence for mostly never-before-seen firing patterns \citep{gupta2010hippocampal}. Therefore, there is a line of evidence from Tetris-studies to sudden wake-ups to dream-lag effects showing that partial or loosely-similar dreams can be triggered most reliably by recently-learned tasks, and yet such inculcated dreams generally take the form of never-before-seen experiences or sequences with the traditional dream-like properties of sparseness and hallucination, matching no specific memory but rather a seeming exploration of the state-space of the task itself. 

It is worth noting that in most cases the sparse, hallucinatory, and narrative properties of dreams are unaccounted for by the consolidation hypothesis. Most dreams do not involve specific memories at all, making the integration of new memories a questionable purpose for dreaming. Indeed, it is openly admitted that the consolidation hypothesis still views dreams themselves as epiphenomena \citep{wamsley2014dreaming}. As we will see, much of the supporting results for the integration, replay, or storage of memories actually fits better with the OBH (Section \ref{sec:neuro_evidence}). 

\subsection{\label{sec:forgetting}Dreams are for selective forgetting}

Notably, Francis Crick and his co-author proposed an alternative purpose for dreams in 1983, which they called "reverse learning" \citep{crick1983function, crick1995rem}. In this hypothesis the point of dreaming is somehow to remove "undesirable" connections and help the brain "unlearn." Yet this hypothesis has been largely ignored in contemporary dreaming research. Instead, the alternative hypothesis, that dreams involve replay or consolidation of memory, became favored by the community \citep{vertes2004memory} due to the excitement around early replay results \citep{wilson1994reactivation}. Contemporary neuroscientific research often views there as being both a consolidation phase as well as a forgetting phase for memories, although this is again predominately associated with slow wave sleep, rather than dreaming specifically \citep{feld2017sculpting}. 

Recently there has been computational work in spiking neural networks showing that in a specific model "reverse learning" can indeed prove helpful. Specifically, authors showed that an anti-learning rule during a "sleep phase" of the network, in the form of anti-Hebbian learning, could break up attractor states that were detrimental to learning \citep{thiele2017wake}. It's worth noting that breaking up detrimental attractor states could be thought of as a form of the OBH; however, in general any sort of "reverse learning" is actually unnecessary from the persepective of the OBH. This is because "reverse learning" approaches differ significantly by focusing on how specific memories are destroyed (via anti-learning mechanisms like a hypothetical "reverse STDP") rather than how corrupted inputs or top-down noise can improve generalization like in the OBH.

\subsection{\label{sec:InSim}Dreams are preparations for real-world problems}

The close correlation between creativity and dreams, as well as the similarity of dreams to simulations or virtual realities, has lead to hypotheses that dreaming can be used to solve relevant real-world problems for the animal. Perhaps the most direct statement of this is the idea that dreams act as rehearsals for stereotypical behaviors for animals in a form of "genetic programming" \citep{jouvet1998paradoxical}. In more contemporary studies this has been referred to as the hypothesis that dreams act as proto-conscious states to prepare for activities during waking behavior \citep{hobson2009rem}.

Similar examples of this hypothesis include an interpretation based off of robots that used simulations to figure out self-models \citep{bongard2006resilient} or that an animal might "dream up strategies for success" at night, like for how to best climb obstacles, like rocks, that it faced during the day \citep{adami2006robots}. Evidence for this sort of hypothesis is that there is a form of neural "pre-play" wherein the sequences of activity predict future behavior of the animal \citep{dragoi2011preplay}. However, in a similar manner to the replay results, it is likely that most of the time pre-play is not the actual specific future sequences of activity, and that most sequences are never instantiated during wake.

Perhaps the most general of these hypotheses of dreams as simulations for solving real-world problems is the idea that dreams are actually for refining the ability to create simulations. This "InSim" hypothesis, which specifically is a hypothesis about the dreams of young children, posits that dreams are chances to create simulations and then test their predictions against the real world upon waking \citep{thill2011inception, svensson2012should}. However, this only applies to young children (with the assumption that the few studies are correct that children's dreams are more "boring" than adults), since, as the authors themselves point out, adult dreams would be consistently invalidated by daily events. Indeed, the phenomenology of dreams as sparse and hallucinatory and fabulist make it unlikely that strategies or abilities or preparations that originate in dreams work at all in the real world.  

However, these types of theories are likely right to view dreams as simulations. Yet according to the OBH the purpose of these simulations is not to refine a particular ability or strategy or plan of action, which is what simulations normally are for. Instead the purpose is to provide "out-of-distribution" simulations specifically to prevent overfitting and improve generalization, wherein overfitting is essentially an unavoidable issue brought about by daily learning and therefore a constant threat to the brain's performance on various tasks.

\subsection{\label{sec:predictive_processing}Dreams benefit predictive processing}

Predictive processing is the view that the brain continuously tries to predict its own future states in relation to extrinsic sensory input \citep{clark2013whatever, hohwy2013predictive}. In a predictive processing framework, predictions traverse in a top-down fashion, while sensory input occurs in a bottom-up fashion. Predictions are then compared with inputs, with the goal of minimizing prediction errors, which corresponds to improving the brain's predictions about its own future states. Prediction error can be minimized by changes in action and behavior, a process called "active inference" \citep{friston2011action}.

Predictive processing has become a popular lens through which to view cognitive science \citep{keller2018predictive}. At the same time, it is controversial in that it is unclear that the cortex actually functions primarily to generate predictions about its own future state \citep{litwin2020unification}, and there are more fundamental criticisms about whether minimizing prediction error via actions (the "dark room" problem) is actually a theoretically coherent goal for a universal principle \citep{sun2020dark}. 

Multiple views of dreams in the light of predictive processing has been put forward \citep{bucci2017sleep} arguing for specific neurophysiological correlates and mechanisms. In general, in such theories the phenomenology of dreams is accounted for by the breakdown of the perception-action loop during dreaming: essentially the brain's activity is dominated by its "priors" rather than bottom-up input. Some predictive processing proponents have put forward views that the role of dreams is not actually to test inferences about actions, but to improve a hypothesized generative model (in this context a model used by brain to make predictions), specifically by reducing this generative model's complexity by pruning redundant synapses \citep{hobson2012waking}. According to this hypothesis, the proposed evolved purpose of REM sleep is to minimize the free energy of the brain. Free energy is essentially the complexity of the brain's model minus the accuracy of the brain's predictions about its own states, and therefore reducing model complexity is important for minimizing free energy. Specifically, the hypothesis put forward in \citep{hobson2012waking} is that synaptic pruning (of the kind proposed in SHY during slow wave sleep \citep{tononi2006sleep}) can help minimize model complexity (and therefore help minimize the free energy) since overall parameters of a model are reduced when synapses are pruned. However, SHY actually specifies that there is a net reduction of synaptic weight specifically so that waking activity is generally unaffected by this change, and additionally it is also specifically associated with NREM and slow wave sleep rather than REM and dreaming.

In \citep{hobson2014virtual} a further related hypothesis without reference to the synaptic pruning of SHY was introduced based on the idea that, according to the free energy principle, the brain is continuously trying to better predict its own future states. Therefore when the brain lacks bottom-up sensory input (i.e., during dreaming) the brain is still compelled to minimize free energy, meaning that the model complexity must implicitly be reduced since any prediction errors can stem only from internal consistency of the model rather than inconsistencies with the outside world. According to this view ". . . we propose that REM sleep is an occasion for reiterating and optimizing a generative model of the embodied self with reference to waking experience." \citep{hobson2014virtual}. However, in general predictive processing and particularly active inference approaches to dreaming face a problem: in the vast majority of dreams agency is actually minimized, not maximized (a view supported by the decreased contribution of prefrontal areas during dreaming \citep{braun1997regional, nofzinger1997forebrain}). That is, dreams are not "causal playgrounds" wherein the outcomes of actions are continuously tested against the perceptions they generate. This may be the case of lucid dreaming, but lucid dreaming accounts for a small minority of dreams \citep{schredl2004lucid}. 

Outside of relying on the assumption that the goal of the brain is always to minimize free energy and that actions are consequential in dreams in the same way they are in wake, another particular concern about the hypothesis is that dreams, given their phenomenology as fabulist and hallucinatory, do not seem very good candidates for minimizing the surprise of predictions created by a gestalt generative model (let alone a model of the "embodied self"). For instance, if dreams were about creating optimal prior beliefs or improving a model's self-consistency by minimizing input-less prediction errors (as in \citep{hobson2014virtual}), the consequence would be that dreams should become less surprising throughout development as input-less prediction error is minimized. This process should lead to the generative model becoming more internally consistent, more parsimonious, tame, and therefore more like the waking world, both over development and over a given night's sleep. Yet this is precisely the opposite of what empirical data show, wherein dreams of children are self-reported as static and uninteresting and become more interesting and surprising across development \citep{foulkes2009children}. Indeed, there is no evidence that dreams become optimized for a lack of surprise or for internal consistency over time. While the world may become more predictable over a lifetime, dreams do not.

However, abstracting away from the specifics of these hypotheses, it should be stressed that the background advantage posited by this free energy approach to dreaming is its effect on model complexity, which is is similar in to the focus of the OBH, since reductions in model complexity in machine learning are generally associated with a reduction in overfitting. Within the free energy principle approach to dreaming this is accomplished by the minimization of errors in a generative model's predictions of itself without inputs, rather than via how corrupted or stochastic inputs assist generalization as in the OBH. This provides strong evidence that the concerns of the OBH can be motivated by diverse takes on brain function. Indeed, the OBH could be thought of as a generalization of the issue model complexity plays in predictive processing to all of learning and performance instead of a just for a particular self-model, and without relying on an assumed drive to minimization prediction errors as the mechanism of action to improve generalization, and the OBH does not share the consequences of this drive like unsurprising dreams. However, the OBH can be motivated entirely independently by common practices in deep learning, their resemblance to dream phenomenology, and the similar challenges the brain faces during its daily learning to those of training a neural network.

\section{\label{sec:overfitting}The overfitted brain hypothesis}

As technology advances science often appropriates new technologies for metaphors that help understand complex systems \citep{lunteren2016clocks}. This has been particularly true of neuroscience \citep{daugman2001brain}. In the past decade it has become apparent that there are many lessons for neuroscience to be taken from brain-inspired deep neural nets (DNNs), which offer a different framework for thinking about learning than standard computer architectures. DNNs are far and away the only successful analog to human intelligence on complex tasks, and they tend to develop brain-like connectivity and representational properties, like grid-cells, shape-tuning, and visual illusions \citep{richards2019deep}. One of the most significant differences between DNNs and the brain is that updating of synaptic weights in accordance with the backpropagation of errors has traditionally been looked on as biologically unrealistic. Yet new research reveals that the brain may implement core features of backpropagation, with several viable candidates such as node perturbations or neural gradient representation by activity differences \citep{lillicrap2020backpropagation}.

Therefore there is good reason for neuroscience to look to deep learning for inspiration, since both are systems that perform complex tasks via the updating of weights within an astronomically large parameter space. It is clear that the challenges the brain and DNNs face during learning and performance on complex tasks overlap significantly. Notably, one of the most ubiquitous challenges DNNs face is a trade-off between generalization and memorization, wherein as they learn to fit one particular data set, they can become less generalizable to others. This \textit{overfitting} is identifiable when performance on the training set begins to differentiate from performance on the testing set. An omnipresent problem within the deep learning community, solutions to overfitting in DNNs most often comes in the form of a noise injection, such as making input data sets corrupted and therefore less self-similar \citep{maaten2013learning}. Perhaps the most common explicit technique to prevent overfitting is dropout, which is mathematically the injection of noise and the corruption of input during learning \citep{srivastava2014dropout}. Notably the more self-similar or biased your sampling of training data is, the more overfitting will be an issue. 

The brain faces these challenges as it learns, since what an organism experiences every day can be highly self-similar and biased in its sampling of the environment. The OBH states that dreams offer a biologically-realistic "noise injection." Specifically, there is good evidence that dreams are based on the stochastic percolation of signals through the hierarchical structure of the cortex, activating the default-mode network \citep{domhoff2015dreaming}. Note that there is growing evidence that most of these signals originate in a top-down manner \citep{nir2010dreaming}, meaning that the "corrupted inputs" will bear statistical similarities to the models and representations of the brain. In other words, they are derived from a stochastic exploration of the hierarchical structure of the brain. This leads to the kind structured hallucinations that are common during dreams. 

The hallucinogenic, category-breaking, and fabulist quality of dreams means they are extremely different from the "training set" of the animal, i.e., their daily experiences. The diurnal cycle of fitting to tasks during the day, and avoiding overfitting at night via a semi-random walk of experiences, may be viewed as a kind of "simulated annealing" \citep{kirkpatrick1983optimization} in the brain. That is, it is the very strangeness of dreams in their divergence from waking experience that gives them their biological function.

To sum up: the OBH conceptualizes dreams as a form of purposefully corrupted input, likely derived from noise injected into the hierarchical structure of the brain, causing feedback to generate warped or "corrupted" sensory input. The overall evolved purpose of this stochastic activity is to prevent overfitting. This overfitting may be within a particular module or task such a specific brain region or network, and may also involve generalization to out-of-distribution (unseen) novel stimuli. As will be discussed, the OBH fits well with the disparate known data about dreams, such as their physiological origin in the form of noise that creates "corrupted features" via neuromodulatory influences, their role in learning, and their importance for problem solving and creativity. However, most importantly, it does not consider dreams as epiphenomena generated by some background process, and it also accounts for, and is motivated by, the actual phenomenology of dreams themselves. The \textit{sparseness} of dreams comes from the "dropout" of bottom-up inputs since they are driven solely by feedback activity, their \textit{hallucinatory} nature comes from the higher-up stochastic origin which means they are purposeful corrupted or warped away from the daily "training set" the organism normally experiences, and their \textit{narrative} nature from the top-down genesis of dreams since the brain understands reality in the form of events and stories. That is, according to the OBH, the distinct phenomenology of dreams exists to maximize their effectiveness at improving generalization and combating mere memorization of an organisms day. The evidence for the OBH, as well as more details about its distinguishing claims, are overviewed in the following section.

\subsection{\label{sec:neuro_evidence}Evidence from neuroscience}

What is the evidence for the OBH from traditional methods of neuroscience? It is worth focusing not on all the studies available, but those that distinguish the OBH from the theory that dreams are correlated (in some unspecified way) with learning.

In human behavioral experiments there is good supporting evidence for the OBH specifically. First, the most effective means of triggering dreams that contain partial similarities to real-life events is through repetitive over-training on a task. Examples of this include extensive playing of games like Tetris \citep{stickgold2000replaying} or ski-simulators \citep{wamsley2010dreaming}, which led to dreams involving the learned task, although not specific repetitions or replays of memories. Put another way: the surest way to trigger dreams about a real-world event is to perform a task repetitively during the day, preferably one that is novel. This creates the condition of the brain being overfitted to the task, which then triggers nightly dreams attempting to generalize performance on the task. Evidence of dreaming about tasks specifically improving daily performance on those tasks can be found for things like mirror tracing \citep{schredl2010sleep} and reading with inverted goggles on \citep{de1996vertical}. Even driving cars seems correlated to dreams about driving \citep{schredl2003continuity}. It is likely therefore there is a homeostatic component to the OBH wherein different modules, processes, or systems within the brain become overfitted from usage, which are then most likely to trigger dreams involving those modules. If so, neurons involved in recent learning would be most affected by overfitting. In this way the OBH can account for many of the statistical "replay" results since those neurons that saw synaptic changes in response learning are most affected by the regularization of dreaming. But the OBH further explains why exact sequence replication is rare and most "replay" is actually never-before-seen firing patterns \citep{gupta2010hippocampal}. 

Another line of evidence for the OBH is that in humans there is evidence of task-dependency when it comes to whether sleep improves learning. For adult humans perceptual tasks showed little to no learning increase from sleep wherein cognitive tasks showed significant gains from learning \citep{doyon2009contribution}. Since it is likely that adult humans already have well-fitted perceptual models, we should expect complex cognitive tasks to trigger more gain from an increase in generalizability. While dream reports are actually less common in young children, particularly below the age of 7, from what can be gleaned children's dreams are much more static and perceptual, focusing on individual scenes rather than full narratives or events, indicating that perceptual systems are likely still being reorganized during dreams \citep{foulkes2009children}. Meanwhile newborns exhibit "active sleep," their version of REM, for 50\% of their 16-18 hours of daily sleep, perhaps indicating that early perceptual models are in constant danger of overfitting.

The OBH is also supported by evidence that sleep does not simply improve memory directly, but affects some aspects of memory more than others. For instance, in a word association test, direct associations, the equivalent of pure memorization, did not benefit that much from a night's sleep, while word associations were better able to resist interfering associations \citep{ellenbogen2006interfering}. This indicates again that memorization is least affected by sleep, but generalized performance is most affected. This holds true even in babies, wherein sleep is correlated with increased generalization and abstraction abilities \citep{friedrich2015generalization, gomez2006naps}.

Additionally, there is evidence from behavioral studies that over-training on a texture-discrimination task leads to decreased performance on it, and that sleep specifically, above and beyond the passage of time, rescues this performance 
\citep{mednick2002restorative}. This fits with anecdotal reports of plateauing in terms of performance on a task, like a video game, only to sleep and have increased performance the next day. 

There is also the long-standing traditional association between dreams and creativity, a rich literature. Anecdotal reports about dreams and creativity are supported by careful studies of how sleep improves abstraction and reasoning on tasks \citep{wagner2004sleep, cai2009rem}. This fits directly with the OBH, since an increase in generalization would directly lead to more insights in complex problems, or better performance on cognitive tasks that require creativity. Indeed, it explains the link between creativity and dreaming better than the hypotheses that dreaming is for the integration of new with old memories, the replay of memories, or their storage.

Finally, it might be argued that it is a problem for the OBH that dreams are generally amnesiac, with explicit memory a rarity during dreams. Would it not be strange then that the content of dreams have any effect on the abilities of a neural network? Here, an important distinction should be made between accessing explicit declarative memories and the general fact that learning involves changes to synapses. For instance, the amnesiac effect during sleep may due to prefrontal inhibition. Just as prefrontal cortex inhibition means that dreams are not recognizable as dreams when they are occurring, it may be that the same inhibition makes it difficult to recall in the sense of cognitive access \citep{siclari2016sleep}. Anecdotal evidence from those with dream journals suggests that paying attention to dreams makes them easier to remember, lending credence to this hypothesis \citep{robb2018we}. Additionally, sudden-waking experiments show that dream content is common and recallable. Likely the effects of not being able to either form or access episodic memories of all dreams at the end of the night are due to the neuromodulatory milieu during sleep.

Supporting the OBH is evidence indicating that dreams lead to synaptic changes in the connectivity of the brain, albeit likely this is not as strongly as waking experience, with episodic memory storage significantly reduced. What is the evidence that synapses change during sleep? Proponents of SHY have argued that there is evidence that synapses change during sleep in the form of synaptic homeostasis, regularization which occurs every night in the form of universal down-scaling of synaptic strength \citep{bushey2011sleep}. However, this has been challenged by the observation of potentiation during sleep \citep{durkin2016sleep}. In general it appears that whether there is net potentiation or depression during waking depends on the task \citep{fisher2016stereotypic}, indicating that learning involves synaptic plasticity in both directions in both wake and sleep \citep{raven2018role}. If during dreams synapses are indeed still plastic, then dreams can leave a synaptic trace that can affect performance.

\subsection{\label{sec:comp_evidence}Evidence from deep learning}

One of the most significant, and ubiquitous, challenges any deep neural network faces is the ability to generalize beyond the data set it has been trained on, that is, to avoid simply memorizing the data set. There has been significant effort in the past decade by the deep learning community to develop methods and techniques to avoid overfitting on particular data sets and, at the broadest level, to allow for extrapolation to never-before-seen data sets. This section overviews three commonly used such techniques within deep learning (and research into artificial neural networks generally). Notably, each embodies some phenomenological property of dreams.

First, there is the method of dropout, perhaps the most widely-used technique for preventing overfitting in deep learning \citep{achille2018information}. Dropout occurs during the training of a network, when inputs are made \textit{sparse} by randomly "dropping out" some of them, a form of regularization during learning which is mathematically similar to a noise injection \citep{srivastava2014dropout}. It is important to note that dreams resemble dropout in their spareness, as they do not contain as much perceptual information, or details in general, as waking experiences. This likely increases the salience of relevant features while minimizing irrelevant features, assisting in generalization by making representations more robust and invariant. Dreams are a lesser or weakened state of conscious experience because of this dropping out of bottom-up stimuli, lacking much of the detail of waking conscious experiences, which, according to the OBH actually assists, rather than hinders, their function.

Second, there is the method of domain randomization used in training deep neural networks. In domain randomization, the inputs during learning are "randomized" in the sense of being warped or corrupted in particular ways. This can drastically assist with learning and generalization. Paradoxically, simulating \textit{hallucinatory} inputs rather than learning off of real inputs helps deep neural networks learn real-world tasks  \citep{tobin2017domain}. Domain randomization has have been used in cutting-edge techniques in deep learning, such as being necessary for having a DNN solve a Rubik's Cube using a robot hand \citep{akkaya2019solving}. Domain randomization resembles the hallucinatory quality of dreams in that dreams depart significantly from normal experiences, as if they have been randomly drawn from a varied set of different domains. 

Of course, the implementation of these techniques must be different for the brain. This is because the brain faces many challenges that artificial neural networks do not. Any organism that implemented dropout or domain randomization during its daily learning would face serious survival issues. Therefore, in order to increase generality and avoid overfitting and pure memorization of the waking data set a dedicated offline period is needed. Sleep, possibly having originally evolved for other housekeeping reasons, is the perfect time.

The third common practice in deep learning that has ties to dreaming is the use of fully simulated data via some generative model. In this context, a "generative model" is when a neural network is trained on data from a domain to output generated data that looks as if it came from the domain but that the network itself generates. Generative models lie behind the success of generative adversarial networks (GANs) and other techniques that allow for cutting-edge performance on complex tasks using sets of feedforward networks that anticipate the other's output \citep{goodfellow2014generative}. It is worth noting that GANs and others often produce notoriously dream-like fabulist outputs \citep{hertzmann2019aesthetics}. Indeed, recently an external generative model that created "dream-like" input helped train a DNN to produce the code behind a given mathematical mapping \citep{ellis2020dreamcoder}. It should be noted that in all these cases the generative model exists outside the network itself, which is not biologically realistic in the case of the brain.

But what about cases where the network itself acts the generative model? In networks that are not purely feedfoward or have external models that can be manipulated by experimenters, the stimulation of higher layers (generally through the injection of noise) can lead to patterns of activity in the lower layers that recapitulate the statistical properties of inputs, as if the network were being stimulated from the bottom-up from imaginary sources. This is likely the case in the brain, wherein stochastic activity high in the hierarchy of brain regions creates hallucinatory patterns of inputs via feedback connections. It is worth noting that the proposal of a "wake/sleep" specific algorithm for unsupervised learning of generative models based on feedback from stochastic stimulation goes back 25 years \citep{hinton1995wake}. While there have even been suggestions that the purpose of sleep involves the "wake/sleep" algorithm itself \citep{sejnowski1995neural}, this algorithm is only for a form of unsupervised learning that creates a generative model and requires several assumptions that are not biologically realistic, like only training one set of connections at a time. It is also clear the brain does much more than just learn a single generative model, and many of the same criticism of predictive processing approaches apply (see Section \ref{sec:predictive_processing}).

However, it is likely the case that dreams are indeed a result of noise in the brain's hierarchical structure which traverses its feedback connectivity, which fits with the evidence that dreams are "top-down" \citep{foulkes2014bottom}. This further fits with evidence that dreaming drawn from the brain's model of the world becomes more narrative and complex over time, particularly during adolescence \citep{strauch2005rem}. By adulthood, dreams take on the \textit{narrative} structure of human cognition wherein stories and metaphors and events make up the core function of thought \citep{lakoff2008metaphors}. Since narratives are the way by which human brains understand the world \citep{lakoff2008women}, stochastic stimulation of the hierarchical structure of the brain produces narratives, which act as hallucinatory and sparse bottom-up input for learning, thus combating overfitting and improving generalization. In this way they are a direct expansion of the normal "training set" of an animal, since narratives and events are how conscious perception itself proceeds and understands the world \citep{james2007principles}.

Overall, while there is no one exact method in deep learning that matches precisely with the OBH, this is likely due to the fact that biological instantiations are always different than their artificial counterparts. The overlap between the phenomenology of dreams and common methods in the field of deep learning for mitigating overfitting, avoiding pure memorization, and assisting generalization lend credence to the idea that the evolved function of dreaming is for precisely these purposes.

\section{\label{sec:predictions}Predictions}

The OBH puts together several lines of investigation under one roof by being explicit about asking how generalization during learning can benefit from dreams. This involves understanding how dreams can help overcome an organism's reliance on memorizing just a day's events, which is often highly statistically biased. The theory makes a number of specific predictions which can be pursued both experimentally as well as theoretically.

\subsection{\label{sec:experiments}Experimental validation}

Experimental investigation of the OBH within neuroscience can consist of several components. Under the OBH, much of the benefit of dreaming is in the realm of generalizability not memorization per se, and this can be differentiated with well-designed behavioral tests. For example, it may be that direct measurement of overfitting is possible in humans. This may include the training of subjects on overly repetitious tasks in order to ensure the condition of overfitting. It may also include using similar techniques as those within deep learning to test for generalization of performance.

In animal models there has not been explicit attempts to separate out the difference between pure memorization and generalization, and the effect of sleep deprivation on each. According to the OBH, memorization should be less affected by sleep deprivation than generalization. Therefore, using mouse models of things like context fear generalization \citep{keiser2017sex} could be examined under conditions of sleep deprivation or, if possible, dream deprivation. Beyond behavioral predictions and subsequent studies, there is also the possibility of attempting to track synaptic plasticity in response to dreams. This may include things like tracking changes in dendritic spine morphology during REM, such have been used to to track spine morphology changes during sleep to investigate SHY \citep{cirelli2020effects}.

Notably, the sort of cognitive flexibility and generalization the OBH claims is the purpose of dreams is highly important for workers and those in the armed forces who sometimes operate under sleep deprivation during critical periods, which can lead to increased accident rates \citep{powell2010sleep}, and has a significant monetary annual impact \citep{leger1994cost}. If it is true that sleep-deprived brains are overfitted, they will be prone to make errors in stereotypical ways. Thus it may be easier to know what types of mistakes will be made by individuals operating in sleep-deprived states and in response build more robust fail-safes. 

Furthermore, the OBH predicts there may be the possibility of \textit{dream substitutions}: dream-like stimuli that are artificially generated to have the properties of dreams, and therefore have a similar ameliorative effect on overfitting. Such dream substitutions, delivered via VR or even video, might provide a simple yet effective means for delaying some of sleep deprivation's cognitive defects. The impact of substitutions can be examined both behaviorally but also at the neurophysiological level of REM rebound \citep{ocampo2000homeostasis}. 

\subsection{\label{sec:computation}Theoretical validation}

The OBH has consequences not just for neuroscience, but also for the field of deep learning. This is particularly true of biologically-realistic models, like large-scale thalamocortical spiking-neuron models, which have previously been used to investigate the development of cortical connectivity and its effect on slow waves \citep{hoel2016synaptic}. It is likely that biologically-realistic spiking neurons which are still trainable in the manner of DNNs \citep{hazan2018bindsnet} can be used to explore the benefits of dreams directly. In such a cortical model, neuromodulation can be used to be intersperse training with periods that mimic sleep stages, cycling first through the real input of its training set, and then hallucinatory corrupted input generated from its top-down connections. This should prevent or delay overfitting. 

What sort of stochastic biases allow for dreams to warp input data in a way that is most efficient for avoiding overfitting? If the OBH is correct, then the sparse and hallucinatory nature of dreams suggest that we should expect warping of input distributions that successfully combats overfitting has these qualities. Such distributions should be \textit{sparse} in that they have less entries than normal inputs, and \textit{hallucinatory} in that they should be clustered in a different way compared to the standard "daily" input. This can be directly tested in state-of-the-art DNNs as well as more biologically realistic artificial neural networks. One potential hypothesis is that a reduction to the learning rate (given the neuromodulatory milieu of dreams) during dream-like input could be especially beneficial for learning in DNNS.

\section{\label{sec:relevancy}Discussion}

The Overfitted Brain Hypothesis (OBH) posits the evolved purpose of dreams is to assist generalization by stochastic corruptions of normal sensory input, which combats the highly biased nature of inputs during an animal's daily learning that can lead to overfitting, a ubiquitous problem in artificial neural networks and machine learning in general. It is supported by both empirical evidence (Section \ref{sec:neuro_evidence}) and theoretical evidence (Section \ref{sec:comp_evidence}). In many cases it can explain observed results better than other hypotheses (comparative hypotheses are discussed in Section \ref{sec:current_theories_REM}). For example, it seems the most effective way to trigger dreams about something is to have subjects perform on a novel task like Tetris repetitiously \citep{stickgold2001sleep, wamsley2010dreaming}, likely because the visual system has became overfitted to the task. Additionally, the OBH explains why "replay" results more often contain never-before-see patterns of activity than actual replays of waking sequences \citep{gupta2010hippocampal}. In another example, it explains the fact that synaptic potentiation occurs during sleep \citep{durkin2016sleep}, indicating that learning during dreams themselves leaves behind a synaptic trace.

The OBH does not necessarily contradict other hypotheses about sleep, for instance, the idea that during certain periods of sleep there is ongoing metabolic waste clearance \citep{xie2013sleep}. In this sense then the OBH speculates that dreaming evolved as an exaptation, wherein sleep evolved for molecular housekeeping purposes and only when brains had to significantly learn during the organism's lifetime did the goal of avoiding overfitting and increasing generalization become adaptive. The OBH does not even contradict some hypotheses about dreams, instead adding new dimensions. Examples of this includes the hypothesis that dreams are a test-bed for strengthening the brain's ability to generate mental imagery during wake, explaining the complexification of dreams from childhood to adulthood \citep{foulkes2009children}. Another example of a novel hypothesis that does not stand in opposition to the OBH is that dreams are for defending neural real estate \citep{eagleman2020defensive} (although this may be contradicted by non-wake activation profiles during dreaming). As a hypothesis it shares similar background concerns with that of the free energy approach to dreaming \citep{hobson2014virtual}, although without assuming that dreams are for testing the predictions of a generative self-model's priors or that dreams should become less surprising over time as input-less prediction error is minimized. Overall the OBH should be viewed as flexible and umbrella hypothesis with many antecedents; after all, it is merely the formal proposal that the corrupted, warped, and stochastic nature of sensory input in dreams serves to improve performance on the brain's daily tasks, motivated by the phenomenology of dreams and common practices in deep learning.

It is also worth noting that within the OBH dream's role in improving generalization may include related things like combating "catastrophic forgetting," which is a problem faced by DNNs that try to train on multiple tasks, and which can also be prevented or alleviated by methods like dropout or more complicated techniques like elastic weight consolidation \citep{kirkpatrick2017overcoming}. Recent research shows that stimuli created from stimulation of a network's top-down connections (which are, according to the OBH, similar to dreams), can indeed help avoid catastrophic forgetting \citep{raghavan2019generative}. So while issues like catastrophic forgetting (the complete unlearning of a task while learning another) is not well documented in humans, it may be that individual brain modules or networks face some lesser form of it, and dreams can be conceptualized as a form of regularization that may ameliorate several aspects of common learning failures simultaneously. For example, beyond the improvements in both in and out-of-distribution generalization, the stochasticity and spareness evinced in the phenomenology of dreams can also likely improve things like security in response to adversarial attacks and overall computational efficiency, as stochastic stimulation has these effects in DNNs \citep{sabuncu2020intelligence}.   

Perhaps the most distinguishing aspect of the OBH is that it takes the phenomenology of dreams seriously, in that they are sparse, hallucinogenic, and narrative in the sense of containing fabulist and unusual events. The OBH emphasizes that it is precisely because of the departure from waking life that dreams evolved. According to this hypothesis dreams are not epiphenomena, either in the sense of neutral evolution but also in the sense of not being an expression of some other background process, such as patterns of activity and associated experiences merely brought about by some other processing integrating new memories \citep{wamsley2014dreaming}. Rather, the point of dreams is the dreams themselves, since they provide departures away from the statistically-biased input of an animal's daily life, which can assist therefore increase performance. It may seem paradoxical, but a dream of flying may actually help you keep your balance running. The evidence for this possibility comes from common methods in deep learning which improve generalization, such as dropout \citep{srivastava2014dropout}, domain randomization \citep{tobin2017domain}, and the use of input data created by stochastic stimulation of generative models \citep{ellis2020dreamcoder}, which together bear striking similarities to the properties of dreams.

The OBH makes several predictions that are useful for both the field of neuroscience and the field of deep learning. These include predictions on the neurophysiological level, as well as behavioral, and even within the field of deep learning. For instance, the prediction that inputs with dream-like properties, i.e., adhering to dream phenomenology, will assist with overfitting in DNNs. Behaviorally, overfitting might be induced in subjects via through repetitive training on an under-sampled task, and the benefit of dreaming might be directly measured. There is also the possibility of dream substitutions, wherein artificial dream-like stimuli might help improve generalization and therefore performance in sleep-deprived individuals.

Finally, it is worth taking the idea of dream substitutions seriously enough to consider whether fictions, like novels or films, act as artificial dreams, accomplishing at least some of the same function. Within evolutionary psychology, the attempt to ground aspects of human behavior in evolutionary theory, there has been long-standing confusion with regard to human interest in fictions, since on their surface fictions have no utility. They are, after all, explicitly false information. Therefore it has been thought that fictions are either demonstrations of cognitive fitness in order to influence mate choice \citep{hogh2018aesthetic}, or can simply be reduced to the equivalent of "cheesecake" --- gratifying to consume but without benefit. Proponents of this view have even gone so far as to describe the arts as a "pleasure technology" \citep{pinker1997mind}. However, the OBH suggests fictions, and perhaps the arts in general, may actually have an underlying cognitive utility in the form of improving generalization and preventing overfitting, since they act as artificial dreams.

\section{\label{sec:acknowledgments}Acknowledgments}

Thanks to Baran Çimen, Santosh Manicka, and 
Hananel Hazan for their comments and feedback.

\bibliographystyle{plainnat}

\bibliography{biblio}

\end{multicols}
\end{document}